\documentclass[prb,aps,amssymb,twocolumn,showpacs]{revtex4}
\usepackage[dvips]{graphicx}
\def\be{\begin{equation}}
\def\ee{\end{equation}}
\def\bea{\begin{eqnarray}}
\def\eea{\end{eqnarray}}
\def\nn{\nonumber\\}
\def\({\left(}\def\){\right)}

\def\Zbar{\bar{Z}}

\begin{document}

\title{Network models for localisation problems belonging to the
chiral symmetry classes}

\author{Marc Bocquet and J.T. Chalker}
\affiliation{Theoretical Physics, University of Oxford,
1 Keble Road, Oxford OX1 3NP, United Kingdom }
\date{\today}

\begin{abstract}
We consider localisation problems belonging to the chiral
symmetry classes, in which sublattice symmetry is
responsible for singular behaviour at a band centre.
We formulate models which have the relevant symmetries and
which are generalisations of the network model introduced previously
in the context of the integer quantum Hall plateau transition.
We show that the generalisations required can be re-expressed as
corresponding to the introduction of absorption and amplification into either
the original network model, or the variants of it that represent
disordered superconductors. In addition, we demonstrate that
by imposing appropriate constraints on disorder,
a lattice version of the Dirac equation with a random vector potential
can be obtained, as well as new types of critical behaviour.
These models represent a convenient starting point for analytic
discussions and computational studies, and we investigate
in detail a two-dimensional example without time-reversal invariance.
It exhibits both localised and critical phases, and
band-centre singularities in the critical phase
approach more closely in small systems the expected asymptotic form
than in other known realisations of the symmetry class.
\end{abstract}


\maketitle

\section{Introduction}
The classification by symmetry of Hamiltonians for disordered systems
provides an important framework in the study of Anderson localisation.
Three standard symmetry classes are long-established, and represented in
the zero-dimensional limit by the three Wigner-Dyson random matrix ensembles.\cite{Mehta1991}
The existence of additional symmetry classes has been implicit in work
which also has an extensive history, but a complete classification
has been set out only rather recently, by Altland and Zirnbauer.\cite{Zirnbauer1996,Altland1997}
Hamiltonians belonging to one of these additional symmetry classes are characterised
by a discrete symmetry which relates eigenvalues and eigenfunctions in pairs.
This paper is concerned with three examples, known as the chiral symmetry classes.
They can be realised as tight-binding Hamiltonians for systems in which sites can be
divided into
two sublattices and non-zero matrix elements connect only sites from opposite sublattices.
In these terms, the discrete symmetry operation is multiplication of wavefunction
components on one sublattice by a factor of minus one: applied to an eigenstate with
energy $E$, this generates a second eigenstate, with energy $-E$. Three distinct such
symmetry classes
are possible, termed\cite{Altland1997} AIII for systems without time-reversal invariance,
BDI for spinless systems with time-reversal invariance, and CII for systems
having time-reversal but not spin-rotation symmetry.

Early work on disordered Hamiltonians with chiral symmetry followed a variety of
independent directions.
A one-dimensional model of this type was solved exactly by Dyson \cite{Dyson1953},
who found a divergence in the density of states at the band centre, of a form shown by
Ovchinnikov and Erikhman
\cite{Ovchinnikov1977} to be characteristic for the symmetry class in one dimension. A closely related feature
of this and similar models is that,
while states away from the band centre are localised,
both the mean free path and the localisation length
diverge on approaching the band centre, as revealed in early treatments
by Eggarter and Riedinger \cite{Eggarter1978}and by Ziman.\cite{Ziman1982}
Two-dimensional systems with chiral symmetry were studied by Gade and Wegner,
\cite{Gade1991,Gade1993}
using a mapping to a non-linear sigma model: their calculations
for the class AIII with weak disorder indicate a strongly divergent density of states
at the band
centre, also accompanied by a divergence of the localisation length.
Chiral random matrix ensembles, investigated by Nagao and Slevin \cite{Nagao1993}
and by Verbaarschot and Zahed,\cite{Verbaarschot1993} likewise show band centre anomalies
in the density of states, albeit on energy scales comparable
with the level spacing.
And in a separate development, extensive analytical results have been obtained
by Nersesyan {\it et al} \cite{Nersesyan1995}
and by Ludwig {\it et al} \cite{Ludwig1994} for a particular
two-dimensional model with chiral symmetry,
the massless two-dimensional Dirac equation with
random vector potential.

Much subsequent work has helped to establish the range of behaviour possible in these systems.
In one dimensional models, a finite localisation length at zero energy and a modification
of the Dyson singularity in the density of states can be induced \cite{Ovchinnikov1977,Comtet1995,Brouwer1998}
by terms in the Hamiltonian that in the absence of disorder would generate a spectral gap.
For two-dimensional systems, it is recognised that a random vector potential
in the Dirac equation
constitutes a special choice of disorder, and scaling flow from this towards a generic
fixed point, with properties similar to those derived by Gade, has been studied
in a field-theoretic framework \cite{Guruswamy2000}
and by using results on random Gaussian surfaces. \cite{Motrunich2002}
Moreover, while early numerical results \cite{Evangelou1986,Morita1997,Furusaki1999}
were quite different from analytical predictions, careful tuning of model parameters
and study of large systems
has recently brought calculations closer to expected asymptotic behaviour.\cite{Motrunich2002,Ryu2002}
Numerical work has also shown that a finite value for the zero-energy localisation length
can be produced in two-dimensional chiral models, by the same mechanism that is
effective in one dimension.\cite{Motrunich2002}

Our aim in the following is to formulate and study new representatives
of the chiral symmetry classes, in the form of network models.\cite{Chalker1988}
In place of a Hamiltonian,
these use the ideas of scattering theory and may be specified by a
transfer matrix,\cite{Chalker1988} or by a unitary
evolution operator for one step of discretised time.\cite{Klesse1995,Ho1996}
The versions we set out here are connected in two distinct ways
to particular network models without chiral symmetry which have been
studied previously. For the class AIII, these connections are to the U(1) network model
\cite{Chalker1988} investigated in the context of the integer quantum Hall plateau transition,
while for CII and BDI, respectively, they are
to the SU(2) and O(1) network models, which describe plateau transitions in
dirty superconductors.\cite{Kagalovsky1997,Kagalovsky1999,Chalker2002a}
In each case, the model with chiral
symmetry is constructed by coupling two copies of the partner model.
This coupling can be re-expressed
after a transformation as equivalent to introduction of absorption and coherent amplification
in the original models. This equivalence parallels the established link \cite{Simons1998}
between chiral symmetry classes and non-Hermitian random operators.

The remainder of this paper is organised as follows. We show
in Sec.\,\ref{CCNM} how to construct network models
belonging to the chiral symmetry classes, and discuss in Sec.\,\ref{exploring}
some basic aspects,
including the relation to non-Hermitian models,
the continuum limit, and a lattice version of disorder
analogous to a random vector potential in the Dirac equation.
We present results from a numerical study
of a two-dimensional model in the symmetry class AIII  in Sec.\,\ref{numerics}, and
summarise in Sec.\,\ref{summary}.


\section{Construction of network models with chiral symmetry}
\label{CCNM}

In this section we formulate network models for each of the chiral
symmetry classes, giving a detailed treatment
of the chiral unitary class (AIII), and indicating in outline the
equivalent steps for the chiral orthogonal (BDI) and
chiral symplectic (CII) classes. Our strategy is simply to
construct the two-dimensional internal space associated with
chiral symmetry using two related copies of a network model
without that symmetry.
For each symmetry class we focus on two-dimensional
models; models in quasi-one and three dimensions can be constructed in the
same way, starting for example from Eq.\,(\ref{TEO}).

The symmetry of a disordered system may be discussed in terms of a Hamiltonian
$H$, a scattering matrix $S$, or a transfer matrix $T$.
We are concerned with systems which conserve probability density,
so that the Hamiltonian is Hermitian and the scattering matrix is
unitary: $H^{\dagger}=H$ and $S^{\dagger}=S^{-1}$. The 
equivalent condition for the transfer matrix involves the current
operator $J$ and reads $T^{-1}=JT^{\dagger}J$. 
Chiral symmetry is implemented on a two-dimensional internal space
in which the Pauli operator $\Sigma_x$ acts: for the Hamiltonian it is the
requirement that $\Sigma_x H \Sigma_x=-H$.
Taking the scattering matrix to have the symmetry of $e^{i H}$,
this implies that $\Sigma_x S \Sigma_x = S^{-1}$. For the transfer matrix, it appears at first 
that one has a choice, taking flux in the pairs of scattering channels on which $\Sigma_x$ acts
to propagate either in opposite directions or in the same direction. In fact only the former is
tenable. It implies that
the anticommutator $\{J,\Sigma_x\} =0$ and that the chiral symmetry condition is $\Sigma_x T \Sigma_x = T$,
a property preserved under matrix multiplication.
By contrast, the latter choice would lead to the commutator $[\Sigma_x,T]=0$ and the chiral symmetry
condition $\Sigma_x T \Sigma_x = T^{-1}$. Since the last condition is not in general preserved under
matrix multiplication, connection in series of two systems of this kind would generate a sample
without chiral symmetry, and we therefore reject this alternative.
Systems in class AIII have no other relevant discrete
symmetry; those in classes BDI and CII are also invariant
under time-reversal, in the absence and presence of Kramers
degeneracy respectively.
\subsection{Network model for class AIII}

\begin{figure}[h]
\includegraphics[width=7cm]{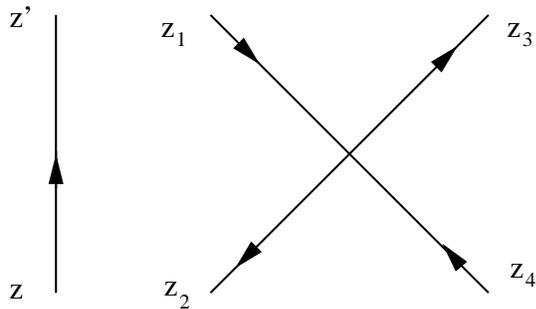}
\caption{\label{building_units}
Units from which the network model is constructed: a link (left)
and a node (right).
}
\end{figure}

\begin{figure}[h]
\includegraphics[width=8cm]{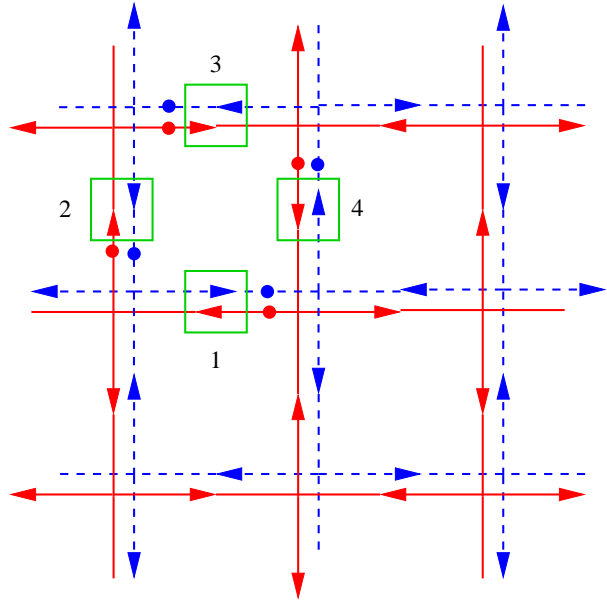}
\caption{\label{fig:network_AIII} Structure of the AIII network model.
Full and dashed lines indicate the two U(1) models from which the system
is constructed, with nodes located at the vertices of the two lattices.
Scattering that couples the sub-systems is represented schematically by
square boxes, shown only in the upper left plaquette.
In one step of time evolution, flux propagates between successive
points marked with filled circles, in the directions indicated with arrows.
}
\end{figure}

We recall first the essential features of the U(1) network
model for the integer quantum Hall plateau transition.
A wavefunction in this model takes complex values $z_l$ on the
links $l$ of the lattice illustrated by the full lines of Fig.\,\ref{fig:network_AIII}.
The forms of the transfer matrix and of the evolution operator follow
from the properties of the elementary building units, shown in
Fig.\,\ref{building_units}. 
A particle acquires a phase $\phi_l$ on traversing link $l$,
so that in a stationary state
amplitudes at either end  are related by
\begin{equation}
\label{link}
z' = e^{i\phi}\,z\,.
\end{equation}
In a similar way, stationary state amplitudes on the four links which meet
at a node are related by a $2\times 2$ transfer matrix
\be
\label{node}
\left( \begin{array}{c} z_1 \\ z_2 \end{array} \right) =
\left(
\begin{array}{cc}
\cosh(a) & \sinh(a) \\
\sinh(a) & \cosh(a)
\end{array}
\right)
\cdot \left( \begin{array}{c} z_3 \\ z_4 \end{array} \right)
\, ,
\ee
where $a$ is real and all phase factors are associated with links.
This equation may re-written in terms of a scattering matrix as
\be
\label{node2}
\left( \begin{array}{c} z_3 \\ z_2 \end{array} \right) =
\left(
\begin{array}{cc}
\cos(\alpha) & -\sin(\alpha) \\
\sin(\alpha) & \cos(\alpha)
\end{array}
\right)
\cdot \left( \begin{array}{c} z_1 \\ z_4 \end{array} \right)
\, ,
\ee
with $\sin(\alpha) = \tanh(a)$.
The transfer matrix that results from assembling
these units is described in detail in Ref.\,\onlinecite{Chalker1988},
and the time evolution operator in Refs.\,\onlinecite{Klesse1995}~and~\onlinecite{Ho1996}.

Introduction of a two-dimensional internal space associated with chiral
symmetry results in a doubling of the number of wavefunction components.
In this way, starting from two copies of the U(1) model the link amplitudes become
two-component complex numbers, ${\bf z}_l$. In place of Eq.\,(\ref{link}),
the scattering properties of a link are characterised by a $2 \times 2$
transfer matrix $T$, with
\begin{equation}
\label{chiral_link}
{\bf z}' = T{\bf z}\,.
\end{equation}
Requiring $\Sigma_x T \Sigma_x = T$, $T$
has the form
\be
\label{link_chiral}
T = e^{i\phi}
\left(
\begin{array}{cc}
\cosh(b) & \sinh(b) \\
\sinh(b) & \cosh(b)
\end{array}
\right) \, ,
\ee
where $\phi$ is a real phase and $b$ is a real hyperbolic angle.
It remains to discuss scattering at nodes of the doubled system.
We replace Eq.\,(\ref{node}) by
\be
\label{node_chiral}
\left( \begin{array}{c} {\bf z}_1 \\ {\bf z}_2 \end{array} \right)
= {\openone} \otimes
\left(
\begin{array}{cc}
\cosh(a) & \sinh(a) \\
\sinh(a) & \cosh(a)
\end{array}
\right)
\left( \begin{array}{c} {\bf z}_3 \\ {\bf z}_4 \end{array} \right)  \, ,
\ee
where $\openone$ is here the unit matrix in the two-component space
introduced on links. This choice amounts to the most general one compatible
with chiral symmetry, since all scattering within the two-component space
may be included in the link transfer matrices, $T$.

Combining these elements to make a two-dimensional system,
we arrive at the model shown schematically in  Fig.\,\ref{fig:network_AIII}.
The transfer matrix for the system as a whole acts in the $[1,1]$ (or $[1,{\overline{1}}]$)
direction and may be written as a product of factors relating amplitudes on successive
slices of the system. Alternate factors in the product represent
links and nodes, and consist respectively of repeated versions of the
$2\times 2$ and $4 \times 4$
blocks appearing in Eqs.\,(\ref{link_chiral}) and (\ref{node_chiral}).
We introduce disorder by taking the phase $\phi$ in
Eq.\,(\ref{link_chiral}) to be an independent,
uniformly distributed random variable on each link, and take the parameter $a$,
characterising scattering at nodes, to be non-random.
We have considered two ways to set the value of the coupling between the
chiral subspaces, which we parameterise in terms of either
the hyperbolic angle appearing in Eq.\,(\ref{link_chiral})
or the compact angle $\beta$ related to $b$ by $\sin(\beta)=-\tanh(b)$:
we take $\beta$ either uniformly distributed, or $b$ to have a normal
distribution of variance $g$.\cite{footnote1}
We ensure that the system is statistically invariant under $\pi/2$
rotations of the lattice, which places requirements on the node parameter $a$,
exactly as in the U(1) model: nodes lie on two distinct
sub-lattices, and the node parameter $a$ on one sublattice is related
to the parameter $a'$ on the other sublattice by
the duality relation $\sinh(a)\sinh(a')=1$.

The same system can alternatively be described using a time evolution operator, $S$,
in place of a transfer matrix. In order to specify this operator,
which is unitary and has the symmetry of
a scattering matrix, it is convenient first to consider the special case $b=0$, in which
the two copies of the network model are uncoupled.
Let $U$ denote the time evolution operator for one copy.
Then, since from Eq.\,(\ref{link_chiral})
link phases are the same in both copies but propagation directions are opposite,
the evolution operator for the other copy is $U^{\dagger}$.
The dimension of $U$ is equal to the number of links in the system.
It is useful to define a diagonal matrix of the same dimension,
with the angles $\beta_l$ for each link $l$ as diagonal entries: we
use $\beta$ to denote this matrix.
The time-evolution operator for the system with chiral symmetry
can then be written
\be
\label{TEO}
S=
\left(
\begin{array}{cc}
U \cos(\beta) & -U \sin(\beta) \, U^{\dagger} \\
\sin(\beta) & \cos(\beta) \, U^{\dagger}
\end{array}
\right) \, .
\ee
It is straightforward to check that $\Sigma_x S \Sigma_x = S^{-1}$ and that
$S^{\dagger}=S^{-1}$.

\subsection{Network model for classes BDI and CII}

A model in class BDI can be obtained from
one in class AIII simply by imposing time reversal invariance as an
additional symmetry. This condition is conventionally
written in the form $H^*=H$, but for a discussion based on
scattering matrices it is more convenient to make the
transformation $H\to QHQ^{-1}$ with $Q=\exp\(i\frac{\pi}{4}\Sigma_x\)$.
This transformation leaves the chiral symmetry relation $\Sigma_xH\Sigma_x=-H$
unchanged. In the transformed basis one has
$H^*=-H$, $S=S^*$ and $T^*=T$.
To ensure a real time evolution operator, we restrict the link phases $\phi$ to
the values $0$ and $\pi$. Choosing these values randomly, the
BDI model consists of two coupled copies of
the class D models studied recently in the context
of disordered superconductors.\cite{Chalker2002a} Alternatively, one could
set $\phi=0$ on all links, and introduce disorder only through the chiral couplings $\beta$.

We turn next to the symmetry class CII. Kramers degeneracy is a defining feature
of the class and necessitates the introduction of an additional two-dimensional
space arising from spin. The time reversal operation includes reversal of spin
direction. Defining ${\cal C} = i \tau_y$, where $\tau_y$
is a Pauli matrix acting in the additional space, it is conventionally
written in the form ${\cal C}H^*{\cal C}^{-1}$=H. As for the class
BDI, it is again convenient to make the transformation 
$H\to QHQ^{-1}$ with $Q=\exp\(i\frac{\pi}{4}\Sigma_x\)$.
In the transformed basis one has ${\cal C}H^*{\cal C}^{-1}=-H$, ${\cal C}S^*{\cal C}^{-1}=S$
and ${\cal C}T^*{\cal C}^{-1}=T$ as equivalent expressions of time-reversal invariance.
Applying these ideas to a network model, four channels
propagate on a single link which, generalising Eq.\,(\ref{chiral_link}),
has a $4 \times 4$ transfer matrix $T$ with the generic form
\be
\label{link_chiral_su2}
T=
v\otimes \left(
\begin{array}{cc}
\cosh(b) & \sinh(b) \\
\sinh(b) & \cosh(b)
\end{array}
\right) \, ,
\ee
where $v$ is an SU(2) matrix and $b$ is a real hyperbolic angle.
Adopting this form, the time evolution operator for class CII has the structure
given in Eq.\,(\ref{TEO}),
but with $U$ representing a class C network model, studied previously
in connection with the spin quantum Hall effect.\cite{Kagalovsky1997,Kagalovsky1999} 
The links of a such a class C model carry two co-propagating
channels, coming from two spin components, and the evolution operator satisfies 
${\cal C}U^*{\cal C}^{-1}=U$.


\section{Discussion of the models}
\label{exploring}

In this section we discuss some basic aspects which are common to
models from all three chiral symmetry classes. These include:
symmetry of the spectrum of the time-evolution operator;
the relation to network models which have absorption and amplification;
the Green function; the continuum limit; behaviour for some special
parameter values; and a lattice
analogue of  random vector potential disorder
in the Dirac equation.

\subsection{Spectral symmetries}
\label{symmetries}

The eigenphases $E_{\alpha}$ and eigenvectors $\psi_{\alpha}$ of the
time evolution operator $S$ play the same role for a network model
as do energy eigenvalues and eigenvectors for a model defined using a Hamiltonian. 
They satisfy
\be
S\psi_{\alpha} = e^{i E_{\alpha}}\psi_{\alpha}\,.
\ee
Chiral symmetry, $\Sigma_x S \Sigma_x = S^{-1}$, has the consequence
that $\Sigma_x \psi_{\alpha}$ is an eigenvector of $S$ with opposite
eigenphase:
\be
S \Sigma_x \psi_{\alpha} = e^{-i E_{\alpha}}\Sigma_x \psi_{\alpha}\,.
\ee
Because of this, the spectrum of eigenphases is symmetric around the point
$E=0$ and their density $\rho(E)$ may develop a singularity
there. In addition, for models from the class CII,
each eigenphase is Kramers degenerate.
Specifically, in addition to $\psi_{\alpha}$, $S$ has a second eigenvector,
${\cal C}\Sigma_x \psi^*_{\alpha}$, with the same eigenvalue 
$e^{i E_{\alpha}}$.

The symmetries described in the preceding paragraph are the ones
of fundamental interest in this work. However, all network models
based on the lattice of Fig.\,\ref{fig:network_AIII} have in addition
a bipartite structure. This structure leads to symmetry
of the eigenphase spectrum under the translation $E\rightarrow \pi + E$.
When combined with chiral symmetry, $E \rightarrow - E$, the consequence is
that behaviour near $E=0$ is mirrored exactly at $E=\pi$, and
that similar behaviour is repeated at $E=\pm \pi/2$.

\begin{figure}[h]
\includegraphics[width=4.6cm]{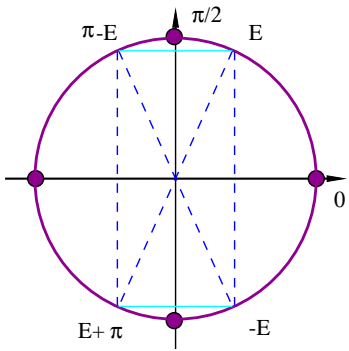} 
\caption{\label{fig:spectre} Symmetries of the
spectrum of the time evolution operator
for a network model defined on the lattice illustrated in
Fig\,\ref{fig:network_AIII}.
Four eigenphases in the spectrum of $S$
are shown as points on the unit circle in the complex plane.
They are related by chiral
symmetry combined with the symmetry of the bipartite lattice.}  
\end{figure}

In detail, the bipartite structure (which is inherited\cite{Ho1996} from
the network model for the quantum Hall plateau transition, and discussed
further in Sec.\,\ref{continuum_limit}) has the
consequence that the evolution operator $U$ appearing in Eq.\,(\ref{TEO})
(or its counterparts for classes BDI and CII) can be written with
a suitable ordering of the basis states as 
\be
\label{bipartite}
U=
\left(
\begin{array}{cc}
0 & A \\
B & 0
\end{array}
\right) \, .
\ee
Introducing in the same basis a diagonal matrix $\sigma = {\rm diag}(+1,-1)$,
bipartite lattice structure implies that $\sigma U \sigma = -U$. Applying this
to our models (and suitably extending the meaning of $\sigma$),
one finds that the pairs of eigenvectors of $S$ identified above, 
$\psi_{\alpha}$ and $\Sigma_x \psi_{\alpha}$, are related to pairs
$\sigma \psi_{\alpha}$ and $\sigma\Sigma_x \psi_{\alpha}$, with eigenvalues
$\pi + E_{\alpha}$ and $\pi - E_{\alpha}$ respectively.
These relationships are illustrated in Fig.\,\ref{fig:spectre}.
They result in an eigenphase spectrum that is reflection-symmetric not only
about the points $E=0$ and $E=\pi$ but also about the points $E=\pm \pi/2$.

\subsection{Chiral symmetry and network models with
absorption and amplification}

Any Hamiltonian $H$ which satisifes the
chiral symmetry condition $\Sigma_x H \Sigma_x= - H$
has a block structure that is most clearly displayed after making the
rotation $R=(\Sigma_x + \Sigma_z)/\sqrt{2}$. In the rotated basis,
the chiral symmetry condition involves $\Sigma_z$ rather than $\Sigma_x$,
since $\Sigma_z = R\Sigma_x R^{-1}$. We have
\be
\label{anti-diagonal}
R H R^{-1}=
\left(
\begin{array}{cc}
0 & {\mathbf h}  \\
{\mathbf h}^{\dagger} & 0 
\end{array}
\right) \,. 
\ee
This establishes a decomposition of the Hermitian operator $H$
into the operators ${\mathbf h}$ and ${\mathbf h}^{\dagger}$,
which are in general non-Hermitian.
We show in this section that for a network model
with chiral symmetry there exists a similar decomposition
into a pair of systems, each with half the number of degrees of freedom.
A single member of the pair, taken in isolation, can be interpreted
as a system in which there is absorption and amplification
of flux. We discuss this separation first in the context of the
transfer matrix, and then in terms of the time evolution operator.

The logic can also be applied in the opposite direction.
In this case, just as Eq.\,(\ref{anti-diagonal}) provides a useful 
relation\cite{hermitisation}
between a non-Hermitian operator ${\mathbf h}$, and the 
Hermitian operator $H$, so
the arguments below can be used to relate a network 
model without flux conservation
to a doubled system in which flux is conserved.

\subsubsection{Transfer matrices with absorption and amplification}

For our models all factors entering the transfer matrix are
diagonal in the two-dimensional space on which $\Sigma_x$ acts,
except for the $2 \times 2$ matrices appearing in
Eqs.\,(\ref{link_chiral}) and (\ref{link_chiral_su2}).
These matrices are diagonalised by the rotation $R$,
which is {\it disorder independent} and which mixes {\it counter-propagating}
channels:
\be
R
\left(
\begin{array}{cc}
\cosh(b) & \sinh(b) \\
\sinh(b) & \cosh(b)
\end{array}
\right)
R^{-1}
=\(
\begin{array}{cc}
e^b & 0 \\
0   & e^{-b}
\end{array} \)
 \, .
\ee
Because this rotation is independent of disorder, we can
apply it to the transfer matrix $T$ for the system as a whole.
The result reads symbolically
\be
\label{splitting_2}
R T R^{-1}= \(
\begin{array}{cc}
{\cal T}(b) & 0 \\
0           & {\cal T}(-b)
\end{array} \) \, .
\ee
In the case of
our model with AIII symmetry, ${\cal T}(b)$ is 
obtained from the transfer matrix
for the corresponding U(1) model by including imaginary parts in
the link phases:
\be
\phi_l \rightarrow \phi_l - i b_l\,.
\ee
In a similar way, for our models with BDI or CII symmetry, ${\cal T}(b)$
is obtained from the transfer matrices for class D or class C models
respectively, by associating an extra factor of $\exp(b_l)$ with
propagation along each link $l$.

If ${\cal T}(b)$ or ${\cal T}(-b)$ is regarded as the 
transfer matrix of a system in its own right, then that system is
one with absorption and amplification. 
In fact, of course, the rotation $R$ gives the current operator $J$
off-diagonal components which couple ${\cal T}(b)$ and ${\cal T}(-b)$,
so that $T$, properly considered, is current-conserving as it should be.

From Eq.\,(\ref{splitting_2}), one sees 
that the spectrum of Lyapunov exponents for a network model
with chiral symmetry consists of separate contributions,
from ${\cal T}(b)$
and from ${\cal T}(-b)$.
Models in which the $b_l$ are random and
distributed symmetrically about $b_l=0$
clearly have doubly degenerate Lyapunov exponents.
These can be calculated by studying just one member of
the pair ${\cal T}(\pm b)$, which is important because it
halves the size of matrices
involved in numerical calculations.

\subsubsection{Time evolution operators with absorption and amplification}

\begin{figure}[h]
\includegraphics[width=6.8cm]{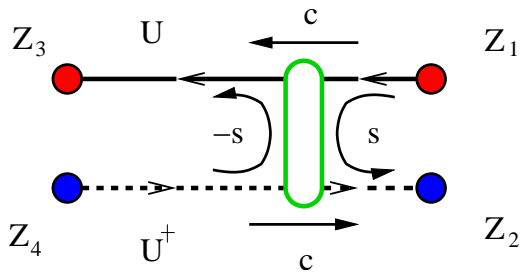} 
\caption{\label{fig:TEO_schema} Schematic representation of the
action of the time evolution
operator $S$ and the matrix $K$,
appearing in Eqs.\,(\ref{TEO}) and (\ref{transfert_chirale}).
Scattering amplitudes within the chiral space 
are denoted by $c=\cos(\beta)$ and $s=\sin(\beta)$. 
}  
\end{figure}

It is interesting to re-examine this separation of the chiral model
into subsystems, starting from the time-evolution operator in place of the
transfer matrix. We begin by writing the action of the time-evolution
operator $S$ in a way that emphasises the $2 \times 2$ block
structure evident in Eq.\,(\ref{TEO}):
\be
\label{S}
\( \begin{array}{c}
Z_3  \\ Z_2
\end{array} \)
=S
\( \begin{array}{c}
Z_1  \\ Z_4
\end{array}  \) \, .
\ee
The dimensions of the vectors $Z$ and $\Zbar$ appearing in Eq.\,(\ref{S})
are the same as those of the matrix $U$ in Eq.\,(\ref{TEO}).
This expression can be reorganised into the form
\be
\label{M}
\( \begin{array}{c}
Z_3  \\ Z_4
\end{array} \)
=K \(
\begin{array}{c}
Z_1  \\ Z_2
\end{array} \) \, ,
\ee
where $K$ is
\be
\label{transfert_chirale}
K=
\left(
\begin{array}{cc}
U & 0 \\
0 & U 
\end{array}
\right)
\cdot
\left(
\begin{array}{cc}
\cosh(b) & \sinh(b) \\
\sinh(b) & \cosh(b)
\end{array}
\right)\,.
\ee
Properly speaking,
since $Z_1$ and $Z_2$ are amplitudes of oppositely propagating fluxes,
$K$ should be interpreted as a transfer matrix, in the way 
illustrated in
Fig.\,\ref{fig:TEO_schema}.
Despite this, our approach is to apply to $K$ the rotation $R$,
and to discuss the rotated matrix as if it were an evolution operator.
Doing this, we find
\be
\label{splitting}
RKR^{-1}=
\left(
\begin{array}{cc}
{\cal U}(b) & 0 \\
0 & {\cal U}(-b)
\end{array}
\right) \, .
\ee
where
\be
\label{calU}
{\cal U}(\pm b) = U \exp(\pm b)\,.
\ee
For the special case $b=0$, the subsystems
represented by full and dashed lines in Fig.\,\ref{fig:network_AIII}
are uncoupled. In this case one has ${\cal U}(0) = U$, which is the
time evolution operator for one subsystem, and its inverse for the
other subsystem. 
For general $b$, one can regard ${\cal U}(\pm b)$ as
time evolution operators for systems with absorption and amplification.
However, it is only the stationary states of these systems
which have a simple physical significance. Such states exist either
for an open system with scattering boundary conditions
(the situation represented by the transfer matrix $T$), 
or for a closed system if a pair of eigenphases are zero
(which can be arranged by fine-tuning system parameters).
In this case, we can define eigenvectors
\be
\label{stat}
Z_{+} = {\cal U}(b) Z_{+} \qquad \mbox{\rm and} \qquad Z_{-} = {\cal U}(-b)  Z_{-}\,.
\ee
Undoing the rotation $R$, the stationary states of $S$
satisfy $\psi_{\pm} = S \psi_{\pm}$, with
\be
\psi_{+}= \( \begin{array}{c} Z_{+} \\ Z_{+} \end{array} \) 
 \qquad \mbox{\rm and} \qquad
\psi_{-}= \( \begin{array}{c} Z_{-} \\ -Z_{-} \end{array} \)\,. 
\ee

\subsection{Green function}

The spectral symmetry of the time evolution operator
for network models with chiral symmetry is also shown
clearly by considering the Green function, defined for complex energy
$z$ by
\be
\label{G}
G(z)=\frac{1}{1-zS} \,\, .
\ee
The combined consequences of the unitarity and chiral symmetry of $S$
together fix the form of $G(z)$, which is particularly simple at $z=1$.
For a discussion of this, it is again convenient to 
make the rotation $R$. 
We set ${\tilde S} = R S R^{-1}$ and 
${\tilde G}(z) = R G(z) R^{-1}$, so that
$\Sigma_z {\tilde S} \Sigma_z = {\tilde S}^{-1}$. 
Unitarity of ${\tilde S}$ implies for ${\tilde G}(z)$ 
\be
{\tilde G}^{\dagger}(z^*)+ {\tilde G}(z^{-1})=1\, ,
\ee
while
chiral symmetry is the requirement that
\be
\Sigma_z {\tilde G}(z) \Sigma_z
+ {\tilde G}(z^{-1})=1\, .
\ee
At $z=1$, these two equations reduce to 
${\tilde G}^\dagger(1)+{\tilde G}(1)=1$ 
and $\Sigma_z{\tilde G}(1)\Sigma_z+{\tilde G}(1)=1$. Their solution fixes
the form for ${\tilde G}(1)$ to be
\be
\label{forme_fonction_Green}
{\tilde G}(1)=\frac{1}{2} \(
\begin{array}{cc}
1 & D \\
-D^\dagger   & 1
\end{array} \) \, ,
\ee
where $D$ is a complex matrix.

Our aim, then, is to determine ${\tilde G}(1)$ 
explicitly for the models 
we have introduced.
We start from Eq.\,(\ref{G}), which we 
write as
\be
G(z) = z S G(z) + 1\,.
\ee
Following the step from Eq.\,(\ref{S})
to Eq.\,(\ref{M}), using the matrix $K$ in place of $S$,
and defining 
\be
{K(z)}=\(
\begin{array}{cc}
1 & 0 \\
0 & z^{-1}
\end{array} \) \,
K \, \(
\begin{array}{cc}
z & 0 \\
0 & 1
\end{array} \) \, ,
\ee
we obtain
\be
G(z)=\frac{1}{2}\left\{ 1 + 
\frac{1 + {K(z)}}{1 - {K(z)} } \Sigma_z \right\} \, .
\ee
This result simplifies further at $z=1$ and
after rotation by $R$
\be
\label{green_function}
{\tilde G}(1)=\frac{1}{2} \(
\begin{array}{cc} \displaystyle{
1} & \displaystyle{\frac{1+Ue^{b}}{1-Ue^{b}}} \\
\displaystyle{\frac{1+Ue^{-b}}{1-Ue^{-b}}}  & \displaystyle{ 1 }
\end{array} \) \, .
\ee
Note that, since 
\be
\left[\frac{1+Ue^{b}}{1-Ue^{b}}\right]^{\dagger}=
-\left[\frac{1+Ue^{-b}}{1-Ue^{-b}}\right]\, ,
\ee 
it is clear the Green function has
the form expected from Eq.\,(\ref{forme_fonction_Green}).

Finally, we remark that similar manipulations may be used to construct a Hermitian operator
which has the same eigenvectors as ${\tilde S}$ and has the block structure 
\be
\label{Hamiltonian}
{\cal H}\equiv \frac{1}{2i} \frac{1-\tilde{S}}{1+\tilde{S}}=
\frac{1}{2i} \(\begin{array}{cc} \displaystyle{
0} & \displaystyle{\frac{1-Ue^{-b}}{1+Ue^{-b}}} \\
\displaystyle{\frac{1-Ue^{b}}{1+Ue^{b}}}  & \displaystyle{ 0 }
\end{array} \) \, .
\ee

\subsection{Continuum limit}
\label{continuum_limit}

The network model representing the quantum Hall plateau transition
is known\cite{Ho1996} to have a continuum limit in which the
time evolution operator is that for a Dirac Hamiltonian in two space 
dimensions, with random
vector and scalar potentials and a uniform mass
which varies through zero on crossing the transition.
In this subsection we investigate the continuum limit for
network models with chiral symmetry. As one might expect, this 
enables us to make links with previous work on the Dirac  
equation with randomness in the classes AIII\cite{Ludwig1994,Nersesyan1995} 
and BDI.\cite{Hatsugai1997} A similar approach should also
be of interest for the CII network model, but we do not
explore it here.

We consider first the model for the class AIII.
We obtain the continuum limit by taking all link phases
and all scattering amplitudes within the chiral space to be small,\cite{footnote2}
so that in Eq.\,(\ref{link_chiral}) $|\phi| \ll 1$ and $|b| \ll 1$,
and by taking the node parameter close to its self-dual value,
so that in Eq.\,(\ref{node2}) $\alpha = \pi/4 + m/2$, with $|m| \ll 1$.

\begin{figure}[h]
\includegraphics[width=7cm]{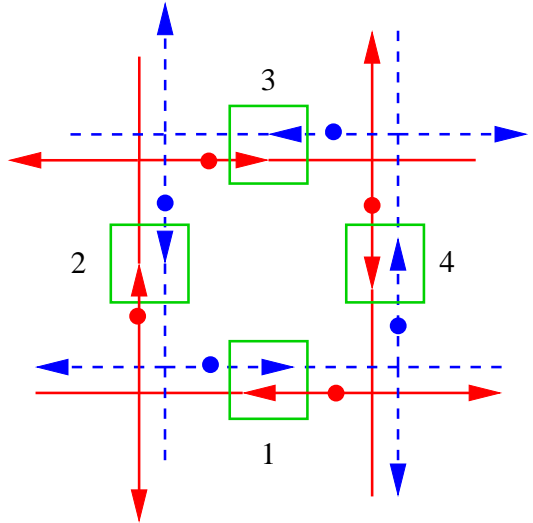} 
\caption{\label{fig:plaquette_AIII_2} A plaquette of the AIII model,
showing in detail the possible scattering processes,
and also the effect of the rotation specified in Eq.\,(\ref{P}). 
Notation is as in Fig.\,\ref{fig:network_AIII}, and the positions of
filled circles should be compared in the two figures.}
\end{figure}

Before going further, it is necessary to deal with a 
technical point which arises because of the bipartite
structure of the models, discussed above in Sec.\,\ref{symmetries}.
This structure has its origin in the form of $U$ displayed in
Eq.\,(\ref{bipartite}), but is obscured in $S$ 
because of the nature of the off-diagonal blocks in Eq.\,(\ref{TEO}).
For that reason, it is useful to make the rotation
represented by
\be
P=
\left(
\begin{array}{cc}
U^{\dagger} & 0  \\
0 & 1 
\end{array}
\right)\,,
\ee
so that the evolution operator is
\be
\label{P}
\hat{S} \equiv PSP^{-1}=
\left(
\begin{array}{cc}
\cos(\beta) & -\sin(\beta)  \\
\sin(\beta) & \cos(\beta) 
\end{array}
\right)
\left(
\begin{array}{cc}
U  & 0 \\
0 & U^{\dagger}
\end{array}
\right)
\, .
\ee
In this basis, the bipartite structure of $U$ is clearly also a
feature of ${\hat S}$. The price paid is that the chiral symmetry 
condition, involving ${\hat \Sigma}_x \equiv P \Sigma_x P^{-1}$,
becomes disorder-dependent (though not at lowest order in $\beta$), 
and for this reason we do not use the basis 
elsewhere in the paper. The effect of the rotation can be
represented by moving some of the points on the network at which
wavefunction amplitudes are defined, as illustrated in 
Fig.\,\ref{fig:plaquette_AIII_2}.

In this basis ${\hat S}^2$, like $U^2$, is
block-diagonal, with separate blocks acting within
each of the two sublattices. 
We obtain a Dirac Hamiltonian $\hat H$ by interpreting the part of 
${\hat S}^2$ that acts within one sublattice as $\exp(-i{\hat H})$,
and expanding to lowest order in $\phi_l$, $b_i \approx \beta_i$ and $m$.
Using the numbering of links around a plaquette specified in 
Fig.\,\ref{fig:plaquette_AIII_2}, we define
\bea
& V=(\phi_1+\phi_2+\phi_3+\phi_4)/2 \, , \nn
& A_x=(\phi_4-\phi_2)/2 \quad {\rm and} \quad A_y=(\phi_1-\phi_3)/2 \,  
\eea
as in Ref.\,\onlinecite{Ho1996}. We also define
\bea
& W=(\beta_1+\beta_2+\beta_3+\beta_4)/2 \, , \nn
& B_x=(\beta_1-\beta_3)/2 \quad {\rm and} \quad B_y=(\beta_4-\beta_2)/2 \, .
\eea
Finally, to put the result in a standard form, we obtain $H$ from
$\hat H$ using a rotation which
takes 
$(\sigma^x,\sigma^y,\sigma^z)$ in $\hat H$ to
$(-\sigma^y,-\sigma^z,\sigma^x)$ in $H$.
We have finally
\bea
\label{climit}
H=&\left[(-i\partial_x-A_x)\sigma^x+(-i\partial_y-A_y)\sigma^y+m\sigma^z+V \right]
\otimes \Sigma_z \nn
&+\left[B_x\sigma^x+B_y\sigma^y+W\right]\otimes \Sigma_y \, .
\eea
As expected, the Hamiltonian $H$ fulfills the chiral 
symmetry condition, $\Sigma_x H \Sigma_x=-H$.
It can be cast in the form of Eq.\,(\ref{anti-diagonal}), with
\be
{\mathbf h}=
(-i\partial_x- {\mathbf A}_x)\sigma^x
+(-i\partial_y-{\mathbf A}_y)\sigma^y+m\sigma^z+{\mathbf V} \, .
\ee
Here, ${\mathbf A}_x=A_x+iB_x$, ${\mathbf A}_y=A_y+iB_y$
and ${\mathbf V}=V+iW$ are, respectively, complex random vector
and scalar potentials.
If disorder is also introduced in the  mass $m$
and as an additional term $B_z\sigma^z\otimes \Sigma_y$,
$H$ is the most general form for a four-component Dirac
Hamiltonian with randomness 
in symmetry class AIII \cite{Guruswamy2000}.

Our network model for class BDI can be treated in a similar
way, but for there to be a continuum limit it is necessary
to restrict link phases to the value zero, instead of allowing
either $\phi_l = 0$ or $\phi_l=\pi$.
We again obtain $H$ in the form of Eq.\,(\ref{anti-diagonal}),
but with
\be
{\mathbf h}=
(-i\partial_x- iB_x)\sigma^x
+(-i\partial_y-iB_y)\sigma^y+m\sigma^z+iW \, ,
\ee
which is a Dirac Hamiltonian with purely
imaginary random vector and scalar
potentials and a real mass $m$.
The same form is obtained from the continuum
limit of the Hatsugai-Wen-Kohmoto model,\cite{Hatsugai1997}
which is known to belong to the symmetry class BDI.

\subsection{Behaviour for special parameter choices}

The behaviour of the models we have introduced can be
simplified by making one of several alternative special choices
for the parameters $\beta_l$ that characterise scattering 
within the two-component space associated with chiral symmetry,
or for the parameter $\alpha$, specifying scattering at nodes.
We discuss these choices in this subsection, restricting ourselves to
the model from class AIII, as illustrated in Fig.\,\ref{fig:network_AIII}.

\subsubsection{No chiral scattering}

On setting $\beta_l=0$ for every link $l$, one has two decoupled
U(1) network models. With the choices for link phases $\phi_l$ and node
parameters $\alpha$ that are set out in Sec.\,\ref{CCNM}, these models
are in localised phases for $\alpha\not= \pi/4$, with a localisation 
length which diverges as $\alpha$ approaches $\pi/4$.

\subsubsection{Uniform chiral scattering}

There are two alternative and natural ways to
pick the parameters $\beta_l$ or $b_l$ without disorder.

One is to set $\beta_l=\beta_0$, a constant. This results in
a model that is statistically invariant under
$\pi/2$ rotations of the lattice. Its time evolution
operator develops a gap in its eigenphase spectrum
near $E=0$ for $\beta_0\not= 0$. To show this, we note
from Eq.\,(\ref{TEO}) that the eigenphases $E_{\alpha}$
of $S$ can be written for constant $\beta_l$ in terms of
$\beta_0$ and those of $U$, which we denote by $e_{\delta}$.
We find
\be
\label{spectre}
\cos (E_\alpha)=\cos(\beta_0) \, \cos(e_\delta) \, ,
\ee 
so that there are no eigenphases $E_{\alpha}$ in the interval
$[-\beta_0, +\beta_0]$. Because of the gap, this choice for
the $\beta_l$ is of limited interest.

An alternative choice is to take the hyperbolic angles constant,
with $b_l = b$ in the transfer matrix. This breaks
the spatial symmetry of the model, because after a $\pi/2$
rotation of the lattice, the sign of $b_l$ is changed in the transfer
matrix for one half of the links of the system. For the transfer
matrix before such a rotation, it is clear from Eq.\,(\ref{splitting_2})
that the Lyapunov exponents of the model with chiral symmetry, 
which we denote by $\nu_n$, are related to those of the U(1) model, $\mu_m$,
by
\be
\label{lyapunov}
\nu_n = \mu_m \pm b\,.
\ee
We take the fact that the Lyapunov exponents of the two models are related
in such a simple way as an indication
that this choice for $b_l$ does not lead to behaviour generic 
for the chiral symmetry class. The result is nevertheless interesting,
because small fluctuations in $b_l$ about an average value $b$
are likely to restore generic properties of the symmetry class
without large changes in the values of the Lyapunov exponents.
In particular, for suitable values of 
$b$ and the node parameter $\alpha$, the spectrum for $\nu_n$ expected
from Eq.\,(\ref{lyapunov}) has a gap around $\nu=0$, indicating
that a localised phase is possible for the network model with chiral symmetry.

\subsubsection{Strong chiral scattering}

\begin{figure}[h]
\includegraphics[width=6cm]{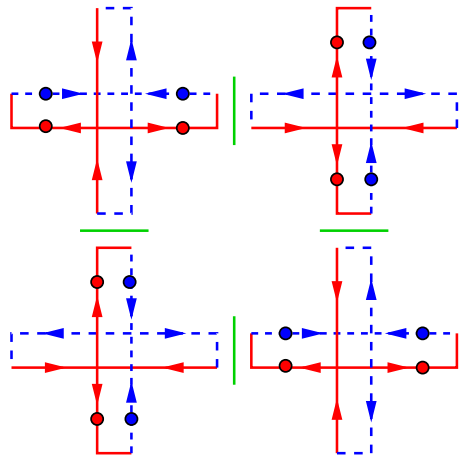} 
\caption{\label{fig:reseau_bipartite} The structure of the AIII network
model in the limiting case $|\beta|=\pi/2$.}  
\end{figure}

The choice $|\beta_l|=\pi/2$ for all $l$ breaks the system up into
a set of disconnected loops, as illustrated in Fig.\,\ref{fig:reseau_bipartite}.
In this limit, possible values for the eigenphases of $S$ are
$E = 0,\pi, \pm \pi/2, , \pm \alpha, \pm \alpha +\pi$ and $\pm \alpha \pm \pi/2$, where
$\alpha$ is the scattering angle at nodes. 
The probability for each eigenvalue to occur depends on the distribution
chosen for the signs of $\beta_l$. To determine the nature
of eigenfunctions as this limit is approached, degenerate
perturbation theory in $|\beta|-\pi/2$ is required.
At leading order, this generates a tight-binding model
with nearest neighbour hopping and site dilution on a square lattice.

\subsubsection{Decoupled plaquettes}

Setting the node parameter to $\alpha=0$ or $\alpha=\pi/2$, the
system breaks up into decoupled plaquettes. Eigenphases for one
such plaquette can be calculated analytically, and vary continuously with
link phases $\phi_l$. Eigenstates are therefore non-degenerate and localised
in this limit.
It is also interesting to consider behaviour at the edge of the system
in a localised phase. 
For a single copy of the U(1) network model which is defined
on the plane rather than a torus, two distinct localised phases arise
in the two limits for $\alpha$: one with current-carrying
edge states and one without. 
Following the construction described in Sec.\,\ref{CCNM}, for models with chiral
symmetry two oppositely propagating edge states appear, with
scattering between them for $\beta_l\not= 0$. Viewed as a strictly
one-dimensional system, these edge states are expected to be critical
if the average $\langle \beta_l \rangle$ is zero, and localised otherwise.
Weak coupling between plaquettes, however, is likely to localise
the edge states for all values of $\langle \beta_l \rangle$.
We therefore do not expect an analogue of the quantum Hall plateau
transition in the models we study, a conclusion that one may also
anticipate from the fact that the models are statistically parity-symmetric.

\subsection{Models with only vector potential disorder}

There has been very extensive interest in Dirac Hamiltonians
on two space dimensions, with only vector potential disorder.
In this subsection we show how lattice versions of such problems can
be realised, using the network model. A similar treatment for
tight-binding Hamiltonians has been described recently by 
Motrunich {\it et al}.\cite{Motrunich2002} 

Our starting point is a special choice for
the chiral scattering angles $\beta_l$, derived from a potential function
as follows.
The potential is defined at the points
which are marked with filled circles on the lattice 
shown in Fig.\,\ref{fig:network_AIII}. 
Suppose that,
on the network drawn with (say) full lines in this figure,
the link $l$ runs from point $i$ to point $j$, and let the potential take
the values $\Phi_i$ and $\Phi_j$ at these points. The 
hyperbolic angle $b$ is then the gradient of the potential, with
\be
b_l = \Phi_j - \Phi_i\,.
\ee
Introducing a diagonal matrix $\Phi$, with $\Phi_l$ as entries, one finds
that the evolution operator ${\cal U}(b)$ appearing in Eq.\,(\ref{calU})
is a similarity transform of the operator $U$
\be
{\cal U}(b) = e^{\Phi} U e^{-\Phi}\,.
\ee
Likewise, one has
\be
{\cal U}(-b) = e^{-\Phi} U e^{\Phi}\,.
\ee

These facts enables us to determine the stationary states for a system
with non-zero $\Phi$ in terms of those for a system with $\Phi=0$, if 
boundary conditions or fine-tuning of the disorder realisation 
allow such states. Specifically, suppose that the system
without chiral scattering has a stationary state $\psi_0$,
so that
\be
U\psi_0 = \psi_0\,.
\ee
Then the system with chiral scattering has stationary states
that, in the notation of Eq.\,(\ref{stat}),
are given by
\be
Z_{+} = e^{\Phi} \psi_0 \qquad 
\mbox{\rm and} \qquad Z_{-} = e^{-\Phi} \psi_0\,.
\ee

A similar simplification applies to the Green function at $z=1$.
The relevant part is the off-diagonal block appearing in Eq.\,(\ref{green_function}),
for which one has
\be
\frac{1+Ue^b}{1-Ue^b}=e^{\Phi} \frac{1+U}{1-U} e^{-\Phi} \, .
\ee
Also the Hamiltonian defined by Eq.(\ref{Hamiltonian}) shows a decomposition reminiscent
of the continuum version
\begin{eqnarray}
{\cal H}_{\Phi} 
&=& 
\left( \begin{array}{cc}  e^{-\Phi}  & 0 \\
0  &  e^{\Phi} 
\end{array} \right)
\left( \begin{array}{cc} 0  & {\cal H}_0 \\
{\cal H}_0  & 0
\end{array} \right)
\left( \begin{array}{cc}  e^{-\Phi} & 0 \\
0  &  e^{\Phi} 
\end{array} \right) \, ,
\end{eqnarray}
where 
\be
{\cal H}_0 = \frac{1}{2i} \frac{1-U}{1+U}\,.
\ee

There are two separate applications of these ideas which are of interest.
The first is to take the operator $U$ to be without disorder, setting all 
link phases $\phi_l=0$. Choosing in addition the node parameter
to have its critical value $\alpha =\pi/4$, $U$ has a stationary state 
$\psi_0$ which is constant in space. Then the wavefunctions $Z_+$ and $Z_-$
are lattice versions of the eigenstates known for the Dirac equation with 
random vector potential, whose statistical properties have been very 
extensively studied. This is supported by the fact that in this case
the continuum limit of the model Eq. (\ref{climit}) simplifies
to a pair of Dirac fermions with real vector potential, equivalent
to the ones described in Ref.\,\onlinecite{Motrunich2002}.

A second application is to include disorder both via $\Phi$ and in $U$ itself.
Taking all link phases to be random, and again setting $\alpha = \pi/4$,
$\psi_0$ is a critical wavefunction for the quantum Hall plateau transition.
The combinations $Z_{\pm} = e^{\pm\Phi}\psi_0$ will therefore
have multifractal fluctuations that are determined from an interplay of the 
distributions of $\psi_0$ and of $\Phi$. It seems very likely
that this will result in a new type of critical behaviour, but we leave 
details for a future investigation.


\section{Numerical results for the two-dimensional AIII model}
\label{numerics}

\subsection{Introduction}

In this section we present results from a numerical investigation of the
two-dimensional AIII network model, as defined in 
Fig.\,\ref{fig:network_AIII} and Eq.\,(\ref{TEO}).
Most of our results are for systems in which the hyperbolic angles $b_l$
have a Gaussian distribution of zero mean and variance $g$. 
For these systems, we examine behaviour as a 
function of the node parameter $\alpha$ and $g$,
parameterising the latter with
$\gamma$,
defined by $\sin(\gamma)=\tanh(\sqrt{g})$.
Hence we consider the parameter space 
spanned by $0\leq \alpha \leq \pi/4$ and $0\leq \gamma \leq \pi/2$.
We also present some data for systems in which the angles $\beta_l$
are uniformly distributed between $0$ and $2\pi$.
To set the scene, we first summarise what can be anticipated for behaviour of the model,
from a consideration of limiting values for the parameters $\alpha$ and $\gamma$ and from
existing analytical results.

We start by recalling that at $\gamma=0$ the system consists
of two decoupled copies of the U(1) network model. States in each 
copy are localised for all $\alpha$ in the range $0\leq \alpha < \pi/4$,
with a localisation length that diverges at $\alpha = \pi/4$.
One expects the localised phase to be stable against small perturbations,
so that it will extend over a region in parameter space with $0\leq \alpha < \pi/4$ and 
$\gamma$ small.
Conversely, critical states in the two U(1) models at $\alpha = \pi/4$
are easily mixed by non-zero chiral coupling $b_l$, so that it is reasonable to guess
that the critical phase predicted by Gade and Wegner\cite{Gade1991,Gade1993} 
will be found in the region with $\alpha$ close to $\pi/4$ and $\gamma > 0$.

Properties along the (critical) line $\alpha=\pi/4$ follow from 
studies of the non-linear sigma model:\cite{Gade1991,Gade1993,Altland1999}
the renormalisation group equations
\be
\label{RG}
\frac{\text{d} \lambda}{ \text{d} l}=0 \qquad
\frac{\text{d} \lambda_A}{ \text{d} l}=\frac{\lambda^2}{16} \, ,
\ee
are argued\cite{Guruswamy2000} to be  exact to all order of perturbation theory.
Note that when $\alpha$ is close to but not exactly equal to $\pi/4$,
a non-zero mean mass for the underlying random Dirac fermions is generated.
In that case, Eq.\,(\ref{RG}) should be extended to incorporate it,
which is beyond the scope of the present work.
This non-zero mass leads to localisation
when large enough on a scale set by $\gamma$.

The coupling constant $\lambda$ parameterises the conductivity $\sigma$,
with $\sigma = 1/(8\pi \lambda^2)$. Since $\lambda$ does not flow under renormalisation,
the critical phase is described by a line of fixed points and
may have a range of values for its conductivity. Flow of the second coupling constant,
$\lambda_A$, determines properties at non-zero energy, giving in particular for
the density of states $\rho(E)$ at small energy $E$ the scaling law\cite{Gade1991,Gade1993} 
\be
\label{GadeSingularity}
\rho(E) \sim \frac{E_0}{E}\exp\(-\kappa \sqrt{\ln\(E_0/E\)}\) \, ,
\ee
where $E_0$ is a microscopic energy scale.
Using the results of Ref.\,\onlinecite{Guruswamy2000}, $\kappa=4\sqrt{2}/\lambda^2$.
As emphasised in Ref.\,\onlinecite{Motrunich2002} such behaviour
for $\rho(E)$ sets in only below a crossover energy scale $E_c$: from
Ref.\,\onlinecite{Guruswamy2000}, we estimate this scale to be given by
\be
E_c=E_0 \exp\(-\alpha\sigma^2\) \, ,
\ee
where $\alpha=(32\sqrt{2}\pi)^2$.
We note also that Motrunich {\it et al} have argued that the
power law in the exponent of Eq.\,(\ref{GadeSingularity}) is
modified at very small energy scales,\cite{Motrunich2002}
and that their results have been reproduced within a field-theoretic approach.\cite{Mudry2002} 

Since the derivations of these results depend on a number of technical assumptions,
some checks are very desirable. Unfortunately, computational work is challenging 
because it is hard to reach scales smaller than $E_c$, and at intermediate 
energies the same analysis predicts an approximately
power-law dependence of the density of states on energy,
with an effective exponent which depends on disorder strength.

In our numerical work testing these expectations, we use calculations of two types,
both applied to quasi-one dimensional systems of width $M$ plaquettes and length  
$L$ plaquettes (so that the time evolution operator $S$ appearing in Eq.\,(\ref{TEO})
is a $4ML \times 4ML$ matrix, and the transfer matrix $T$ 
appearing in Eq.\,(\ref{splitting_2}) is a $4M \times 4M$ matrix). 
For such samples, we find the localisation length $\xi_M$ as the inverse of the
smallest positive Lyapunov exponent for the transfer matrix, calculated using
the standard approach.\cite{Pichard1981}
Applied directly to the transfer matrix as formulated in Sec.\,\ref{CCNM},
the resulting localisation length is that for eigenstates of the time evolution
operator at eigenphase zero; by associating an extra phase $e^{iE}$ with every
link (breaking chiral symmetry), we are also able to find the localisation length
for states with non-zero eigenphase.  
Separately, we determine the density of states for eigenphases of the time evolution
operator $S$, applying a version of the recursion method,\cite{MacKinnon1981} adapted
to treat a unitary operator in place of the Hermitian one for which the
approach was originally formulated. 
We use the terms eigenphase and energy interchangeably in describing our results.

\subsection{Phase diagram}

We determine a phase diagram for the model in the $\alpha$, $\gamma$ plane
from a study of the zero-energy localisation length $\xi_M$ in quasi-one dimensional 
geometry, and its dependence on $M$. (We do not apply this
approach to models with uniformly distributed $\beta_l$, since
our transfer matrix calculations develop numerical
instabilities if $|\cos(\beta_l)| \ll 1$.)
The quantity of central interest is the ratio $\xi_M/M$.
In a localised phase, $\xi_M$ approaches a finite limit $\xi$, the bulk localisation
length, for large $M$, and so the ratio $\xi_M/M$ decreases
with increasing $M$, varying as $\xi/M$ for sufficiently large $M$. 
By contrast, in a critical phase, the ratio approaches 
for large $M$ a finite constant whose value can be identified with
the conductivity of the model.\cite{Chalker1993}

\begin{figure}[h]
\includegraphics[width=7.5cm]{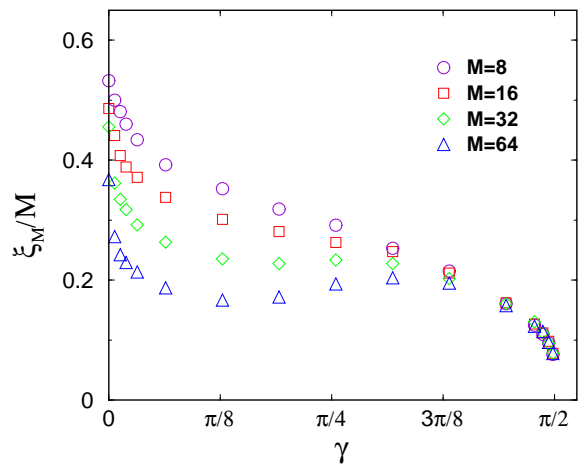} 
\caption{\label{fig:conductivity_1} $\xi_M/M$ at $\alpha=7\pi/32$ as a function
of $\gamma$. }  
\end{figure}

\begin{figure}[h]
\includegraphics[width=7.5cm]{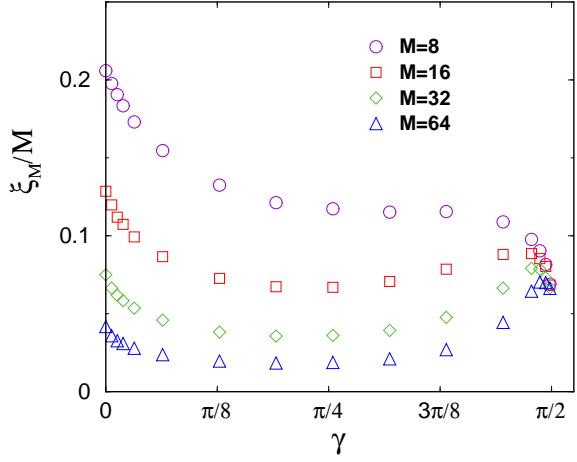} 
\caption{\label{fig:conductivity_2} $\xi_M/M$ at $\alpha=\pi/8$ as a function
of $\gamma$. }  
\end{figure}

\begin{figure}[h]
\includegraphics[width=8cm]{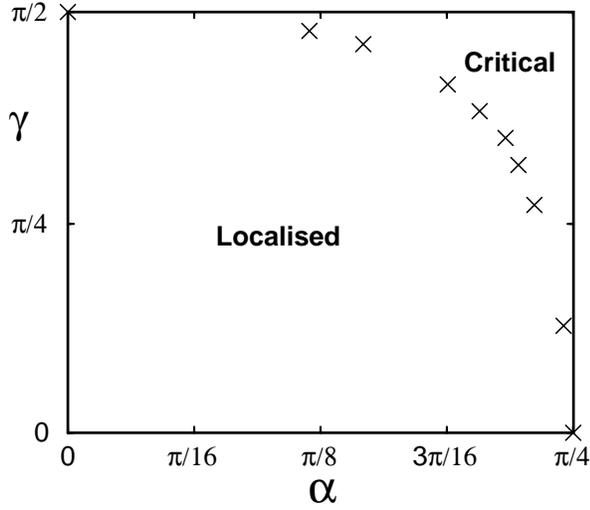} 
\caption{\label{fig:diagram_AIII} Phase diagram of the AIII network model
in the $(\alpha,\gamma)$ plane.}  
\end{figure}

Representative data for $\xi_M/M$ at $M=8, 16, 32$ and $64$ are 
shown as a function of $\gamma$, in Fig.\,\ref{fig:conductivity_1} for
$\alpha=7\pi/32$, and in Fig.\,\ref{fig:conductivity_2} for
$\alpha=\pi/8$.
In both cases, two regimes of behaviour are observed: for small $\gamma$,
$\xi_M/M$ decreases with increasing $M$, indicating a localised phase,
while for $\gamma$ close to $\pi/2$, $\xi_M/M$ is independent of $M$ for the system widths
studied, suggesting a critical phase. By identifying the value of $\gamma$
which divides the two regimes and studying its dependence on $\alpha$,
we arrive at the phase diagram displayed in
Fig.\ref{fig:diagram_AIII}.

\begin{figure}[h]
\includegraphics[width=7.5cm]{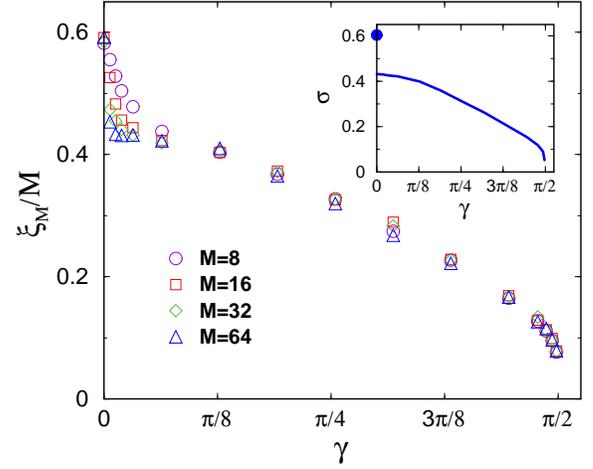} 
\caption{\label{fig:conductivity_3} $\xi_M/M$ at $\alpha=\pi/4$ as a function
of the $\gamma$. Inset: the limiting value, $\sigma$, of this ratio for large $M$}  
\end{figure}

Properties in the critical phase are illustrated in more detail
by the behaviour of the ratio $\xi_M/M$ as a function of $\gamma$ with
$\alpha=\pi/4$, shown in Fig.\,\ref{fig:conductivity_3}. On this line in parameter space,
behaviour at all points
is critical rather than localised, and the data show a finite limiting
value $\sigma$ for the ratio at large $M$, with $\sigma$
dependent on $\gamma$: such behaviour is expected
from Eq.\,(\ref{RG}) and the identification of the coupling constant $\lambda$
as being a function only of conductivity, which is given in turn by $\xi_M/M$.
This figure also illustrates crossover of behaviour with increasing $M$
at small $\gamma$, from that of the U(1) model to that of the AIII model.
The crossover is responsible for discontinuous variation of $\sigma$ with $\gamma$
at $\gamma=0$.

\begin{figure}[h]
\includegraphics[width=7.5cm]{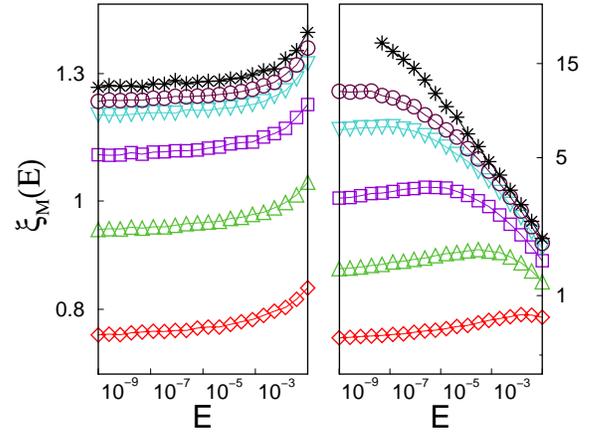} 
\caption{\label{fig:localisation_length_critical} Localisation length $\xi_M$
at $\alpha=\pi/8$, $g=1$ equivalent to $\gamma=0.55 \pi/2$ (left) and at $\alpha=\pi/4$, $g=9$ 
equivalent to $\gamma=0.936 \pi/2$ (right) as a function of $E$
in a log-log scale for increasing transverse size M=2,4,8,16,24 and extrapolated
to $M=\infty$, indicated by the
symbols $\Diamond , \bigtriangleup , \Box , \bigtriangledown , \bigcirc$ and $\star$
respectively.}  
\end{figure}

We show in Fig.\,\ref{fig:localisation_length_critical} the energy dependence
of the localisation length at a point in the localised phase (left panel)
and at a point in the critical phase (right panel), with data at $L=2.5 \times 10^5$
for a range of system widths
$M$, as well as an extrapolation to the
two-dimensional limit. 
It is clear from the data that the localisation length
in the critical phase decreases
very rapidly with $E$ away from $E=0$.

\subsection{Density of states}

\begin{figure}[h]
\includegraphics[width=8cm]{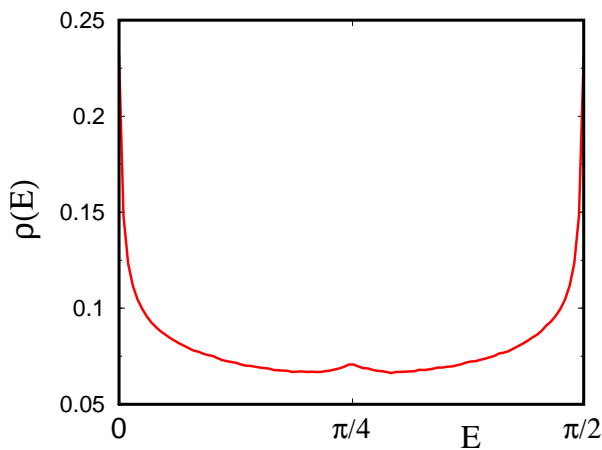} 
\caption{\label{fig:densite_AIII_global} Density $\rho(E)$ of eigenphases $E$
for $\alpha = \pi/4$ and $\beta_l$ uniformly distributed.}  
\end{figure}

Our interest in the (average) eigenphase density of the time evolution operator
is focussed mainly on behaviour near zero energy, where we expect
in the critical phase singular behaviour approaching that of Eq.\,(\ref{GadeSingularity}),
and in the localised phase either a density vanishing quadratically
with energy, as for the random matrix ensemble in this symmetry class,\cite{Zirnbauer1996}
or singularities arising from Griffiths strings, as discussed
in Ref.\,\onlinecite{Motrunich2002}.
Before examining the small energy region in detail, we give an overview of 
behaviour for the entire range of eigenphases, shown 
in Fig.\,\ref{fig:densite_AIII_global} 
for $\alpha=\pi/4$ and uniformly distributed $\beta_l$.
(Similar results are obtained with Gaussian $b_l$ at $g\sim 1$.)
From the discussion of Sec.\, \ref{symmetries}, we need to display data only
over the range $0\leq E \leq \pi/2$.
These results were obtained for a system size of $M=16$ and $L=10^5$,
which is large enough for $\rho(E)$ to be self-averaging at the scale
used here.

\begin{figure}[h]
\includegraphics[width=8cm]{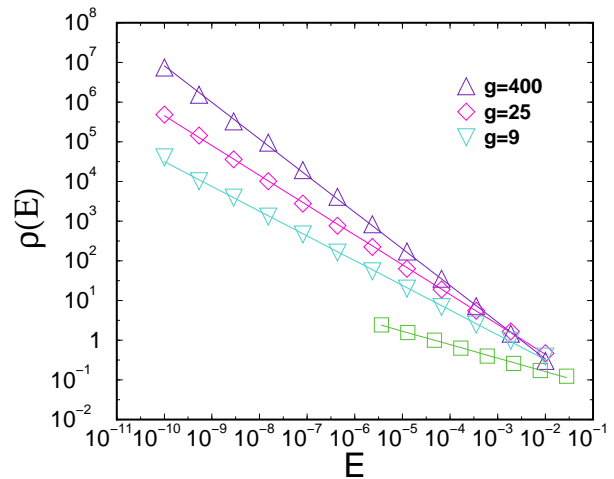} 
\caption{\label{fig:density_AIII_low_E} Density of states $\rho(E)$
as a function of $E$ for small $E$, with logarithmic scales on both axes.
Three cases of increasing disorder are plotted (with symbols $\bigtriangleup$ for $g=400$,
$\Diamond$ for $g=25$ and $\bigtriangledown$ for $g=9$).
The related dynamical exponents $z$ in this range of energy are respectively
$z=26.0 \pm 2.0$, $z=8.0 \pm 0.1$ and $5.3 \pm 0.1$.
Symbols $\Box$ show the case $\beta_l$ uniformly distributed in $[0,2\pi]$
with $z=3.02 \pm 0.03$.}  
\end{figure}

A divergence of $\rho(E)$ as $E \to 0$ is apparent in Fig.\,\ref{fig:densite_AIII_global},
which we now investigate in more detail. In Fig.\ref{fig:density_AIII_low_E} we show  
$\rho(E)$ as a function of $E$, with logarithmic scales for both axes, for a sequence
of models, with $\alpha=\pi/4$ and $b_l$ Gaussian distributed in each case.
Members of the sequence are chosen to have successively smaller values
of $\sigma$, and hence larger values for the energy scale $E_c$, since
by increasing $\gamma$
we decrease $\sigma$, as is evident from the inset to Fig.\,\ref{fig:conductivity_1}.
The data were obtained using systems of width $M=16$, and lengths in the range
$L= 10^5$ to $L=10^9$, in order to achieve satisfactory self-averaging.
Over the energy range accessible, the variation of $\rho(E)$ with $E$
is approximately power-law, and can be characterised following 
Ref. \onlinecite{Motrunich2002} using an effective dynamical exponent $z$, with
\be
\label{power}
\rho(E) \sim |E|^{-1+2/z} \, .
\ee
As one approaches the asymptotic low-energy behaviour, the effective
exponent $z$ increases, ultimately to infinity:
the large values of $z$ reached here indicate close approach
to the limiting low-energy behaviour.

\begin{figure}[h]
\includegraphics[width=8cm]{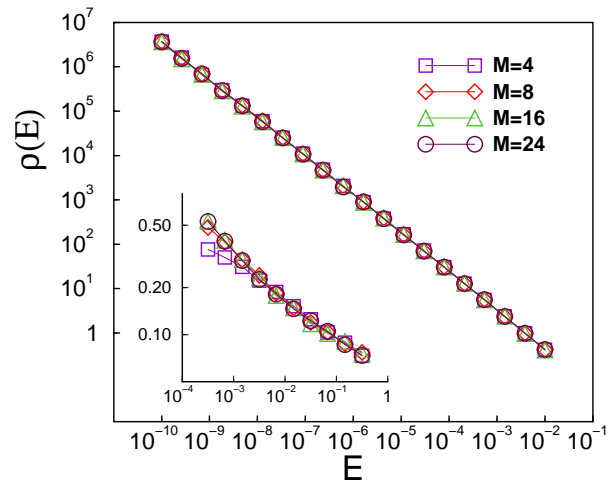} 
\caption{\label{fig:Ly_dependence} Density $\rho(E)$ of eigenphases $E$
for $\alpha = \pi/4$, $g=100$  and for increasing transverse sizes M=4,8,16 and 24.
The inset corresponds to the same set of system sizes but with $\beta_l$ uniformly distributed
in $[0,2\pi]$.}  
\end{figure}

By repeating such calculations for a range of $M$ values, 
we have checked that these results are not influenced by finite system
width system. This is shown in Fig. \ref{fig:Ly_dependence}, in which
the behaviour with uniformly distributed $\beta_l$ is also illustrated.
The rather rapid convergence to the large $M$ limit is
a consequence of the small values of the localisation length for energies away from $E=0$,
illustrated above.

\begin{figure}[h]
  \includegraphics[width=8cm]{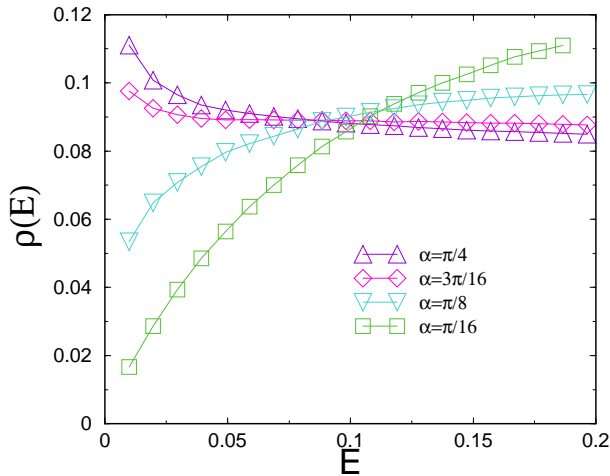} 
\caption{\label{fig:density_AIII_localised} Density of states $\rho(E)$
as a function of $E$ in the range $[0,\pi/16]$ for fixed $g=1$, with linear scales on both axes.
Four cases of decreasing $\alpha$ (moving into the localised phase) are plotted with
$\bigtriangleup$ for $\alpha=\pi/4$ (critical case), $\Diamond$ for $\alpha=3\pi/16$,
$\bigtriangledown$ for $\alpha=\pi/8$ and $\Box$ for $\alpha=\pi/16$.}  
\end{figure}

Finally, we examine the evolution of the
density of states as one moves from the critical phase into the
localised phase. Results are presented in Fig. \ref{fig:density_AIII_localised}.
At the broad scale to which these calculations are restricted,
behaviour in the localised phase is consistent with the power law of Eq.\,(\ref{power}),
with a power $-1+2/z$ which, deep in the localised phase, is positive and
increases as the localisation length decreases. Such behaviour 
was proposed as generic for localised systems with chiral symmetry
in Ref.\onlinecite{Motrunich2002}. 


\section{Summary}
\label{summary}

In summary, we have introduced network models which are realisations
of the chiral symmetry classes. We have argued that they are interesting from
several points of view. They are useful as a starting point for numerical 
studies of these symmetry classes, offering access in the two-dimensional
case we have examined to both critical and localised phases, and
displaying band-centre singularities in the critical
phase which approach quite closely the expected asymptotic form.
The models also have a striking connection to network models  
without chiral symmetry, but with absorption and amplification.
Moreover, by imposing constraints on the disorder, they serve as
lattice versions of problems with randomness entering only
through a vector potential. It seems likely that disorder of this kind
can generate types of critical behaviour different from those known
previously for localisation problems in two dimensions.

\section*{Acknowledgements}      
This work was supported in part by the EPSRC under grant GR/J78327.
We thank N. Regnault for use of computer
facilities.



\begin{references}
\bibitem{footnote1} We have also studied models with $b$ of fixed
magnitude and random sign. Although the asymptotic behaviour of the density of states
appears the same with this choice, strong oscillations appear in an intermediate 
energy regime and make the analysis more difficult.
\bibitem{footnote2} We choose signs in the scattering matrices at nodes
in such a way that, if link phases $\phi_l$ are set to zero, an overall phase factor
of $-1$ is acquired on moving around any elementary plaquette.
\end{references}
\end{document}